# Ghost-Stochastic Resonance in a Unidirectionally Coupled and Small-World Networks

**S. Rajamani\* and S. Rajasekar†**

*School of Physics, Bharathidasan University,*

*Tiruchirapalli 620 024, Tamilnadu, India.*

## Abstract

Ghost-stochastic resonance is a noise-induced resonance at a missing fundamental frequency in the input signal. In this paper we investigate the features of ghost-stochastic resonance in a unidirectionally coupled network and small-world network with each unit being bistable Bellows map. In the one-way coupled network we apply a multi-frequency signal $(1/n_f) \sum_{j=1}^{n_f} \cos(\omega_j n)$, $n = 0,1,2,...$, $\omega_j = (k+j-1)\omega_0$, with $k$ being an integer $\geq 2$ (without the fundamental frequency $\omega_0$) and noise to first unit only. We show the occurrence of resonance and undamped signal propagation for coupling strength above a certain critical value. The response amplitude shows sigmoidal function type variation with unit number. We report the effect of coupling strength $\delta$, $k$ and $n_f$ on the response amplitude. In the small-world network randomness in the connectivity topology is described by the probability $p$ of rewiring of units in a ring type regular network where all the units are subjected to noise and multi-frequency signal. We present the influence of $p$ and the coupling strength on the probability distribution of response amplitude $Q$ of various units, $\langle Q \rangle$ and the maximum value of $Q$.

\*Electronic address: rajeebard@gmail.com

†Electronic address: rajasekar@cnld.bdu.ac.in , srj.bdu@gmail.com

# I INTRODUCTION

One of the fundamental phenomena displayed by dynamical systems is resonance. The term resonance refers to a process in which the response of a system at a particular frequency is maximized when a control parameter is varied. Systems are found to show different kinds of resonance depending upon the nature of the input signal. Linear systems can show frequency resonance [1, 2] a resonance observed by varying the frequency of the applied external periodic force with single frequency, and parametric resonance [3, 4] a resonance due to the periodic variation of a control parameter. Nonlinear systems are capable of showing these two resonances and a few other types of resonance. For example, nonlinear systems exhibit autoresonance [5, 6], stochastic resonance [7, 8], vibrational resonance [9, 10], ghost-stochastic resonance [11, 12], ghost-vibrational resonance [13, 14] and coherence resonance [15, 16]. In autoresonance the frequency of the external driving force is time-dependent and monotonically increases with increase in time. The resonance at the frequency of the external periodic force due to the applied noise is termed as stochastic resonance. In vibrational resonance setup a nonlinear system is subjected to a biharmonic force with frequencies $\omega$ and $\Omega$, $\omega \ll \Omega$. When the amplitude $g$ of the high-frequency periodic force is varied the response amplitude at the frequency of low-frequency signal displays resonance peak at one or more values of amplitude of the high-frequency signal. In certain nonlinear systems where the driving external force consists of multiple frequency with missing fundamental frequency then an additive external noise can induce a resonance at the missing fundamental frequency. This phenomenon is ghost-stochastic resonance. Occurrence of a resonance at the missing fundamental frequency due to a high-frequency external force is called ghost-vibrational resonance. Resonance can be realized in a nonlinear system without any external periodic force but by an external noise source and is the coherence resonance.

The goal of the present paper is to report our investigation on the ghost-stochastic resonance in unidirectionally coupled network and small-world network with each unit being the Bellows map. Because a discrete equation (map) as a unit of a network requires relatively very less computational time and resources compared to a unit being a continuous time dynamical system. In recent years a great deal of interest has been focused on exploring the various nonlinear phenomenon exhibited by networks of unidirectional and small-world structures. Visarath In et al [17-19] introduced one-way coupling to realize oscillatory motions in undriven overdamped and bistable systems. In certain unidirectionally coupled systems propagation of waves of dislocations in equilibria [20], propagation of localized nonlinear waves [21, 22], propagation and anhilation of solitons [23, 24], undamped signal propagation with enhanced response amplitude [25-28], ghost-vibrational resonance [13], oscillatory behaviour [29], spiral and limit cycle dynamics [30, 31], multi stability and a stable limit cycle [33] and quasi-stable travelling wave periodic solution [33] have been found to occur.

There is another class of fascinating network called small-world network [33, 34]. A pioneering report of Watts and Strogatz on a small-world network initiated considerable interest of research activities on the properties of small-world network of different types. For an interesting early experiment of small-world network one may refer to the refs. [35, 36]. Small-world networks are proposed to provide a successful tool to identify connectivity details of anatomical and functional networks in brain [37], and to get deep understanding of complex patterns of activity taking place in neuronal circuitry. The features of ghost resonance has been investigated in a semiconductor laser [38], bidirectionally coupled two lasers [39], vertical cavity surface emitting lasers [40], Schmitt trigger [41], Chua's circuits [42, 43] and $n$- coupled neurons [44] and $n$- coupled Duffing oscillators [13].

The organization of the paper is as follows. In Sec. II we consider a one-way coupled regular network with $N(=2)$ units with first unit alone driven by a multi-frequency force and noise. Each unit is chosen as Bellows map. The external force is $(1/n_f) \sum_{j=1}^{n_f} \cos(\omega_j n), n = 0,1,2,..., \; \omega_j = (k+j-1)\omega_0$ where $k$ is an integer $\geq 2$, $(1/n_f)$ is the number of frequency components in the driving force and $n = 0, 1, 2, ..., n$. First, we report the occurrence of ghost-stochastic resonance in the first unit, the dynamics of which is independent of the dynamics of other units. We characterize the underlying resonance in terms of response amplitude at the missing fundamental frequency $\omega_0$ of the input signal. Next, we explore the effect of number of units in the network, coupling strength and the number of frequency components in the external driving force. Interestingly the response amplitude $Q_i(\omega_0)$ at the missing frequency $\omega_0$ of the input signal display sigmoidal type variation with the unit number $i$ for coupling strength $\delta$ above a critical value for a fixed value of noise intensity. $Q_L = Q_{2000}/Q_1$ the response amplitude of the last unit shows resonance behaviour with noise intensity while it increases with increase in the coupling strength for fixed values of noise intensity. We show the effect of $k$ and $n_f$ on the gain in the response amplitude $Q_g = Q_{2000}/Q_1$. Section III is devoted for the analysis of ghost-stochastic resonance in small-world network where all the units are driven by the external force and noise. The network is Watts-Strogatz with rewiring probability $p$. The emphasis is on the effect of $p$ and the coupling strength on $\langle Q \rangle$ and $\langle Q \rangle_{max}$ (maximum values of $\langle Q \rangle$). Finally in Sec. IV we present the conclusions.

## II  One-Way Coupled Regular Network

We consider a one-way coupled regular network with $N$ units and first unit alone driven by both noise and multi-frequency force. The first unit is uncoupled while the other units are unidirectionally coupled. For simplicity the coupling is chosen as linear and each unit as the Bellows map [24-26]. The iterative coupled maps of our interest are

$$x_{n+1}^{(1)} = \frac{rx_n^{(1)}}{1+\left(x_n^{(1)}\right)^b} + \frac{f}{n_f}\sum_{j=1}^{n_f}\cos\left(\omega_j n\right) + \sqrt{D}\xi(n), \tag{1a}$$

$$x_{n+1}^{(i)} = \frac{rx_n^{(i)}}{1+\left(x_n^{(i)}\right)^b} + \delta x_n^{(i-1)}, \quad i = 2,3,...,N, \tag{1b}$$

Where $\omega_j = (k+j-1)\omega_0$ with $k$ being an integer $\geq 2$, $D$ is the intensity of noise and $\delta$ is the coupling strength. $\xi(n)$ is a Gaussian white noise with zero mean, intensity $D$ and $\langle \xi(n)\xi(n') \rangle = \delta(n-n')$. The single Bellows map $x_{n+1} = rx_n/\left(1+x_n^b\right)$ is monostable with one fixed point $x^* = 0$ for $0 < r \leq 1$. For $r > 1$, it is a bistable system with two stable fixed points $x_{\pm}^* = \pm\sqrt{r-1}$ and an unstable fixed points $x_0^* = 0$. For our numerical investigation of the system (1) we fix $r = 2$, $b = 2$, $f = 0.4$, $\omega_0 = 0.05$, $k = 2$ and $N = 200$. The choice $k = 2$ gives $\omega_j = (k + j - 1)\omega_0$ $n = 0, 1, 2, \ldots, n_f$. The external force has frequencies that are integer multiples of $\omega_0$ except the missing fundamental frequency $\omega_0$. For $f = 0.4$ the simple Bellows map driven by the external periodic force with $n_f$ frequencies is in the subthreshold region, that is, in the absence of noise there is no transition between the two stable fixed points.

We characterize the noise-induced resonance at a frequency $\omega$ by the response amplitude $Q_i$ given by

$$Q_i = \frac{\sqrt{Q_{i,c}^2 + Q_{i,s}^2}}{f}, \tag{2}$$

where

$$Q_{i,c} = \frac{2}{Tt} \sum_{n=1}^{Tt} x_n^{(i)} \cos \omega n, \quad Q_{i,s} = \frac{2}{Tt} \sum_{n=1}^{Tt} x_n^{(i)} \sin \omega n. \tag{3}$$

In Eqs. (3), $T$ is chosen as 1000 and $t$ is the period of the input periodic driving force. $Q$ is often used as a measure for stochastic resonance [45-48].

First, we present the response of the first unit of the network (1), the dynamics of which is independent of the dynamics of the other units. Equation (1) is numerically iterated with an initial value $x_0^{(1)}$. From the solution of (1a) after leaving sufficient transient evolution the response amplitudes $Q_1(\omega)$ at various frequencies are computed as a function of the noise intensity $D$. Figure 1 shows the variation of $Q_1(\omega)$ for $\omega = \omega_0$, $\omega_1$ and $\omega_2$ with the noise intensity $D$. A typical noise-induced resonance is realized at these frequencies. $\omega_1 = 2\omega_0$, and $\omega_2 = 3\omega_0$, are present in the input signal. The resonance observed with these frequencies is the usual stochastic resonance. The resonance associated with the missing frequency is the ghost-stochastic resonance. The ghost-stochastic resonance occurs at $D = 0.033$. At this value of $D$, $x_n^{(1)}$ versus $n$ plot displays almost periodic switching between the regions $x > 0$ and $x < 0$. Erratic switching between $x > 0$ and $x < 0$ occurs for $D \gg 0.033$.

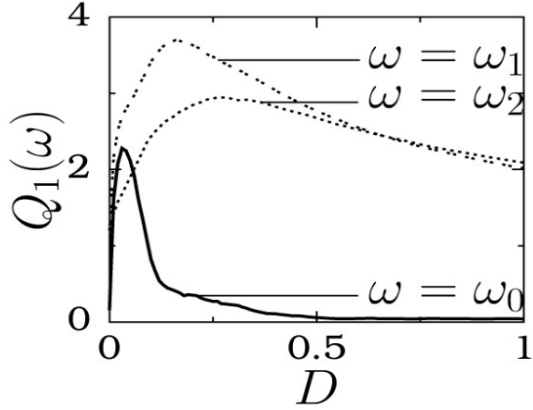

FIG.1. $Q_1(\omega)$ at $\omega = \omega_0$, $\omega = 2\omega_0$ and $\omega = 3\omega_0$ versus the noise intensity $D$ for a single Bellows map. The values of the parameters in Eq. (1a) are $r = 2$, $b = 2$, $f = 0.4$, $\omega_0 = 0.05$, $k = 2$ and $n_{\mathrm{f}} = 2$.

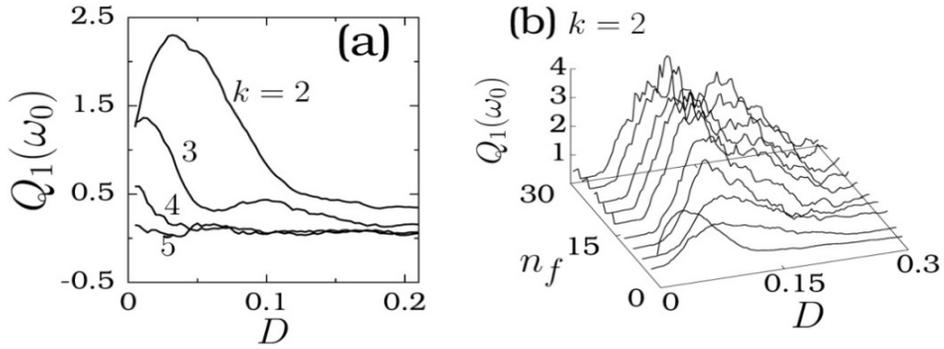

FIG.2. (a) $Q_1(\omega_0)$ as a function of $D$ for and $k = 2, 3, 4$ and $5$ for single Bellows map. (b) $Q_1(\omega_0)$ versus $D$ and $n_{\mathrm{f}}$ for $k = 2$.

Figure 2a describes the effect of $k$ for a few values of it on $Q_1(\omega_0)$. In this figure one can clearly notice resonance for $k = 2$ and $3$. $Q_1(\omega_0)$ at a value of $D$ decrease with increase in $k$ and decays to zero. The value of $D$ at which resonance occurs is reduced when the value of $k$ is increased. The effect of increasing the value of $k$ is to degrade the response of the system at the missing fundamental frequency. A different result is realized when the number of frequencies $n_f$ is increased for $k = 2$. Figure 2b presents the resonance profile for a range of values of $n_f$. A typical resonance occurs for each value of $n_f$. The value of $D_{max}$ and the value of $Q_1$ at $D = D_{max}$ denoted as $Q_{1,max}$ increase and attain a saturation.

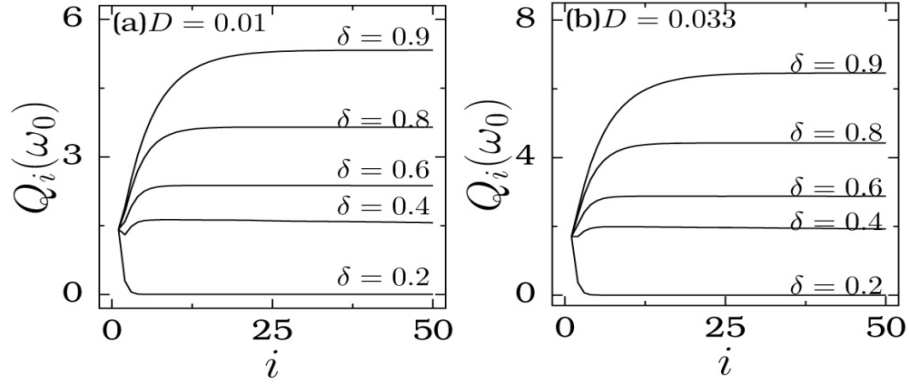

FIG.3. Variation of the response amplitude $Q_i$ at the missing frequency $(\omega_0)$ with the unit number $i$ for five fixed values of $\delta$ with $n_f = 2$ and $k = 2$ and for two values of the noise intensity (a) $D = 0.01$ and (b) $D = 0.03$.

Next, consider the $N$-coupled Bellows maps and focus on the effect of the noise intensity and the coupling strength on the response amplitudes $Q_i$ at the missing frequency $\omega_0$. For each fixed value of $D$ and $\delta$ the response amplitude $Q_i$ monotonically either increases or decreases with $i$ and attains a saturation. For example, Fig. 3 shows the trend in the variation of $Q_i$ with $i$ for two fixed values of $D$ and for five fixed values of $\delta$. For both $D = 0.01$ and 0.03 $Q_i$ decays to zero for $\delta \leq \delta_c = 0.32$ and shows sigmoidal type variation with $i$ for $\delta > \delta_c$.

We numerically computed the limiting value of $Q_i$ in the limit of $i \to \infty$ denoted as $Q_L$ for a wide ranges of $D$ and $\delta$. $Q_L$ shows different kind of dependence on the parameters $D$ and $\delta$. In Fig 4a $Q_L$ displays resonance-like behaviour for the values of $\delta > \delta_c$. For a fixed $\delta < \delta_c = 0.32$, $Q_L \approx 0$. We studied the change in the dynamics of $x_n^{(1)}$ and $x_n^{(10)}$ versus $n$ for $D = 0.004$. $x_n^{(1)}$ switches between the regions $x > 0$ and $x < 0$. The switching is random and also in the regions $x > 0$ and $x < 0$ the iterated values of $x$ oscillates randomly. $x_n^{(i)}$ for $i \gg 1$ becomes rectangular pulse like solution with the width of the pulse being random (see $x_n^{(10)}$ in Fig 5a). Similar behaviour is realized for other fixed values of $D$. The results for $D = 0.033$ and $D = 0.2$ are shown in Figs. 5b and 5c, respectively.

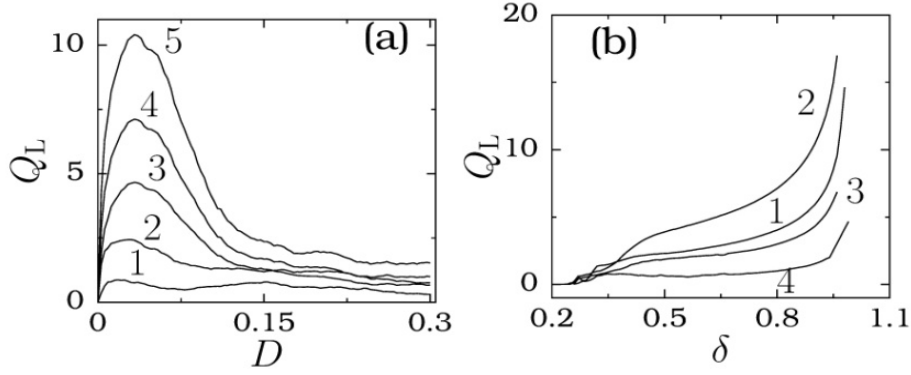

FIG. 4. Resonance-like variation of $Q_L$ with noise intensity $D$ for four fixed values of $\delta$. The values of $\delta$ for the curves 1-4 are 0.4, 0.6, 0.8 and 0.9, respectively. (b) $Q_L$ versus $\delta$ for four fixed values of $D$. The values of $D$ for the curves 1-4 are 0.005, 0.033, 0.1 and 0.3, respectively. In all the cases $n_f = 2$ and $k = 2$ in Eqs. (1).

We define the gain in the response amplitude of the last unit in the coupled systems at the missing frequency $\omega_0$ as

$$Q_g = Q_{2000}/Q_l. \tag{4}$$

$Q_g$ is computed for a certain range of values of $\delta$ and $D$ for $k = 2$ and $n_f = 2, 3, 4$ and $5$. The result is presented in Fig. 6. Colour code representation of $Q_g$ versus $\delta$ and $D$ for $k = 2, 3, 4$ and $5$ is shown in Fig. 7. In Fig. 8 $Q_L$ as a function of $D$ is plotted for $k = 2, 3$ and $6$ and for $\omega = \omega_0$, $\omega_1$, $\omega_2$ and $\omega_3$ with $\delta = 0.7$, $n_f = 2$ and $L = 2000$. For a range of $D$ the value of $Q(\omega_0)$ is not zero. However, $Q$ at $\omega = \omega_1$ and $\omega_2$ are $> Q(\omega_0)$ for a wide range of values of $D$.

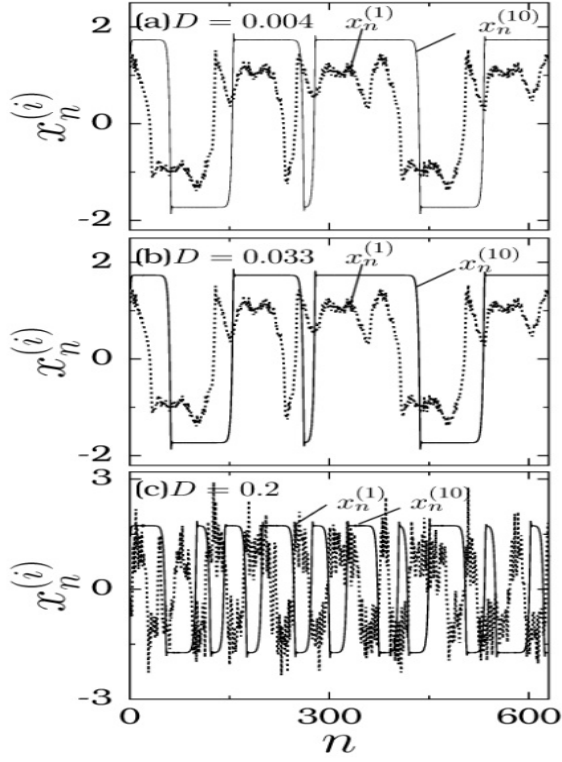

FIG. 5. Time series plot of first and 10th unit of $N$ coupled Bellows maps with $\delta = 0.5$ .

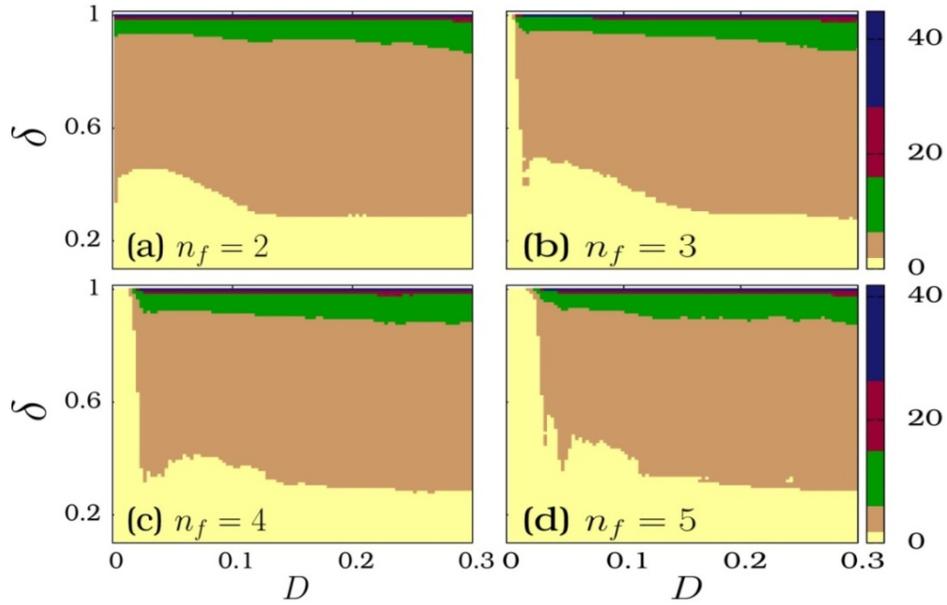

FIG. 6. Colour code representation of $Q_s$ as a function of $\delta$ and $D$ for four values of $n_f$ with $k = 2$ .

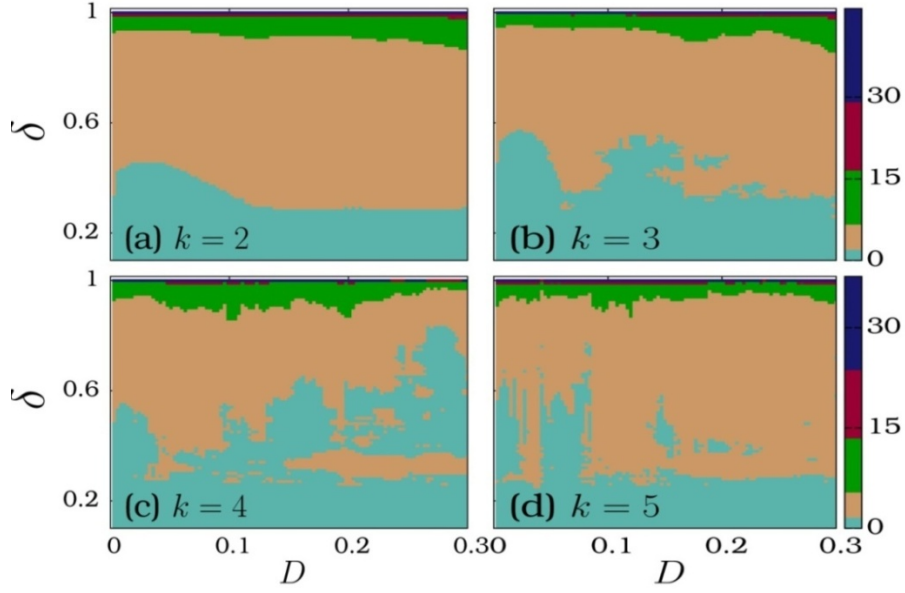

FIG. 7. Same as Fig. 6 but for $n_r = 2$ and for $k = 2, 3, 4$ and $5$.

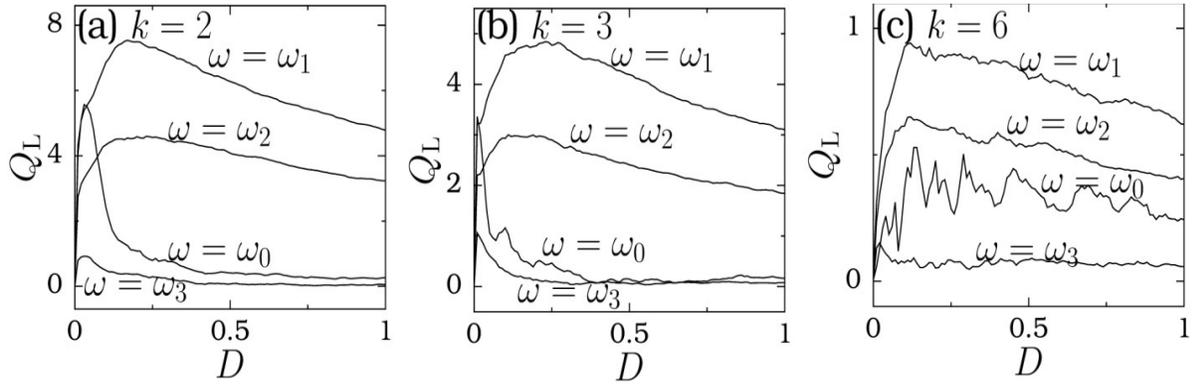

FIG. 8. Variation of $Q_L$ with $D$ for a few values of $k$ and $\omega$ for the $N$-coupled Bellow map.

## III Small-World Network

In this section, we consider an another type of network system, namely, a small-world network which has received a great deal of interest in recent years.

### A. Construction of small-world networks

The construction of a network of a small-world type begins with a simple regular network. For example, consider a circular lattice consisting of $n$ nodes and $k$ edges per node.

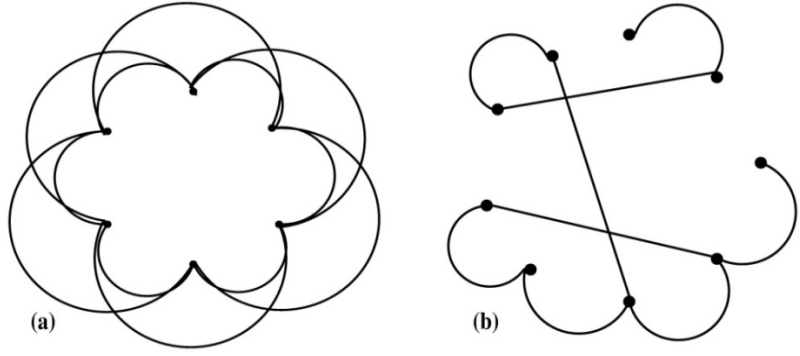

FIG. 9. (a) A regular network with nearest and next nearest neighbouring connectivities. (b) An example of a small-world network.

A typical example of a regular network with 6 nodes and four edges per node is depicted in Fig. 9a. First each node is connected to its nearest and the next nearest neighbour nodes. We note that for each of the $n$ nodes there is a one nearest node to its left and another one to its right. That is, $i$ th node is connected to both $(i-1)$ and $(i+1)$ th nodes. The connection to $(i-2)$ and $(i+2)$ th nodes corresponds to the next nearest neighbour connectivity. A network with the connection to nearest neighbouring nodes alone is a regular network with $(k=2)$ edge. In this way one can construct a regular network with $k$ edges.

To construct the Watts-Strogatz small-world network we rewire each edge at random with probability $p$. Suppose $p=0.5$ and say, there are N nodes. For an $i$ th node we generate a random number between $0$ and $1$ using a uniform random number generator. If the number is $> 0.5$ then we leave the connection between $i$ and $(i+1)$ th nodes. If the number is $\leq 0.5$ then we generate another integer random number between $1$ and N. If this number is $i$ then we generated another number and so on until the generate $j$ number is different from $i$. Suppose the number is $j$. Then we connect the $i$ th node to the $j$ th node. We repeat the above for all the nodes. In the above whenever the generated random numbers is $< 0.5$ the connection between, say, $i$ th node and $(i+1)$ th node is removed and $i$ th node is joined to $j$ th node. This corresponds to $k=2$. The network with $p=0$ is a purely regular one. A network with $p=1$ is purely a random network. For $0 < p < 1$ the network is in between completely regular and completely random. Fig. 9b depicts a small-world network of Watts-Strogatz type with 8 nodes, $k = 2$ and $p = 0.5$.

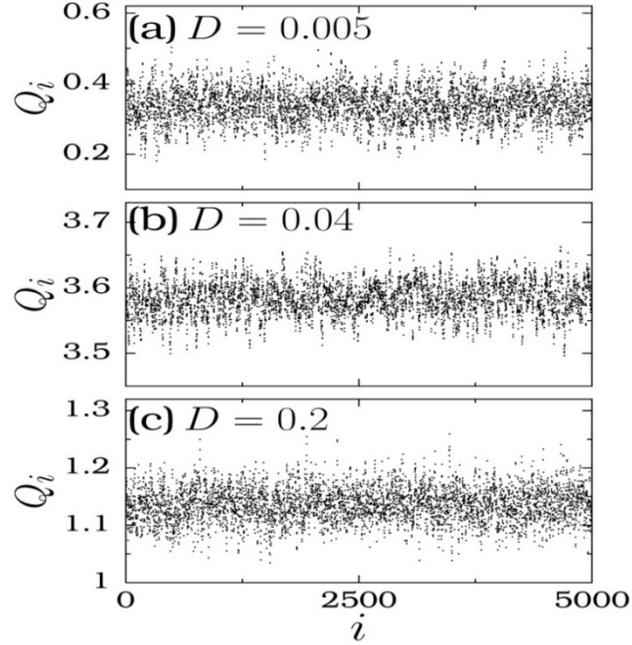

FIG. 10. $Q_i$ versus $i$ for the regular network of Bellows maps for three values of noise intensity with $\delta = 0.5$.

## B. Ghost-Stochastic Resonance

We study the ghost-stochastic resonance in the regular and small-world network systems with each unit being the Bellows map. In our numerical study we fix $r = 2$, $b = 2$, $f = 0.4$, $\omega_0 = 0.05$ and the number of units as $N = 5000$. Figure 10 shows the variation of the response amplitude $Q_i$ at $\omega = \omega_0$ as a function of the unit number for three fixed values of noise intensity with the coupling strength $\delta = 0.5$ and $p = 0$, $k = 2$ and $n_f = 2$. That is, the network is the regular network with bidirectional nearest neighbour coupling. $Q_i$'s are randomly distributed. In Fig. 10 we notice that the mean value of $Q_i$ is large for $D = 0.04$. The result for $p = 0.2$ and $p = 1$ are presented in Figs. 11 and 12, respectively. $p = 0.2$ corresponds to the case in which 20% of the randomly chosen nearest neighbour connections are disconnected and then connected to randomly chosen units. For $p = 0.2$ and 1 also $Q_i$'s are randomly distributed, however, the range of $Q_i$'s varies with $p$.

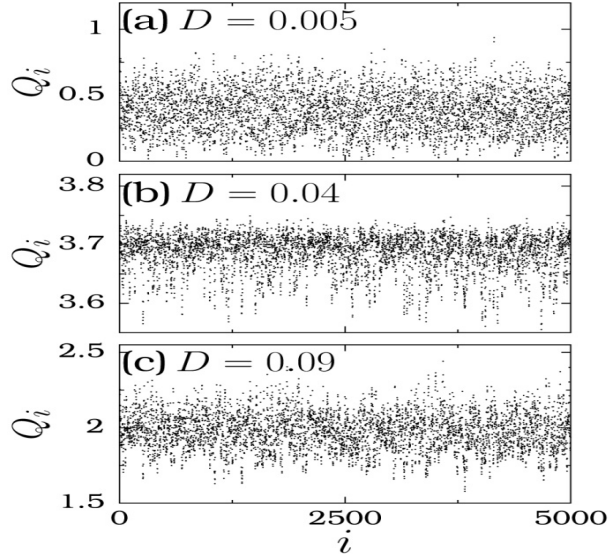

FIG. 11. $Q_i$ versus $i$ for the small-world network of Bellows maps for three values of noise intensity with $p = 0.2$ and $\delta = 0.5$.

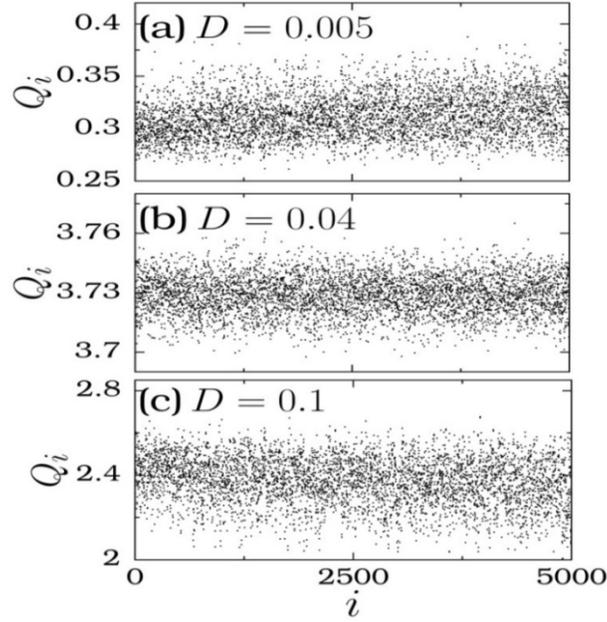

FIG. 12. Variation of $Q_i$ with $i$ for the purely random network of Bellows maps for three fixed values of noise intensity $D$.

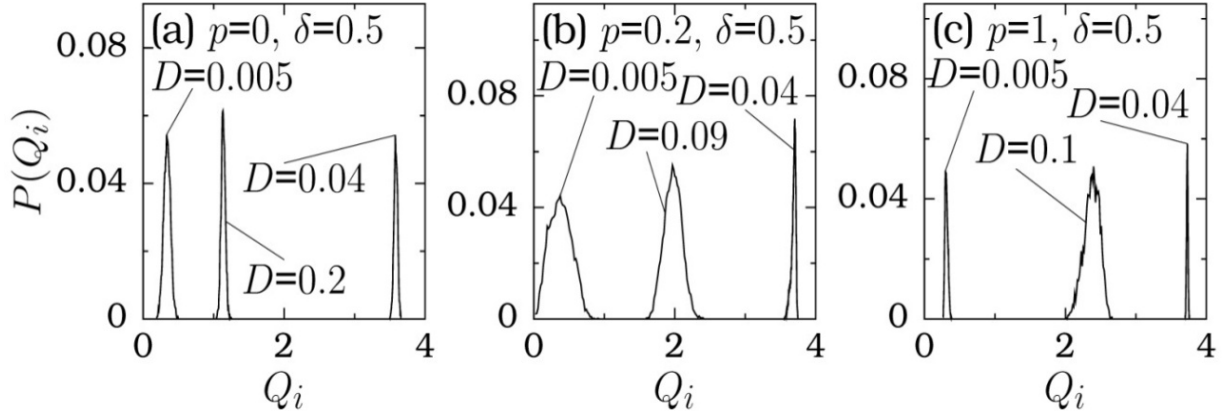

FIG. 13. Plots of $P(Q_i)$ versus $Q_i$ of the network of Bellows maps for three values of $p$ and noise intensity $D$.

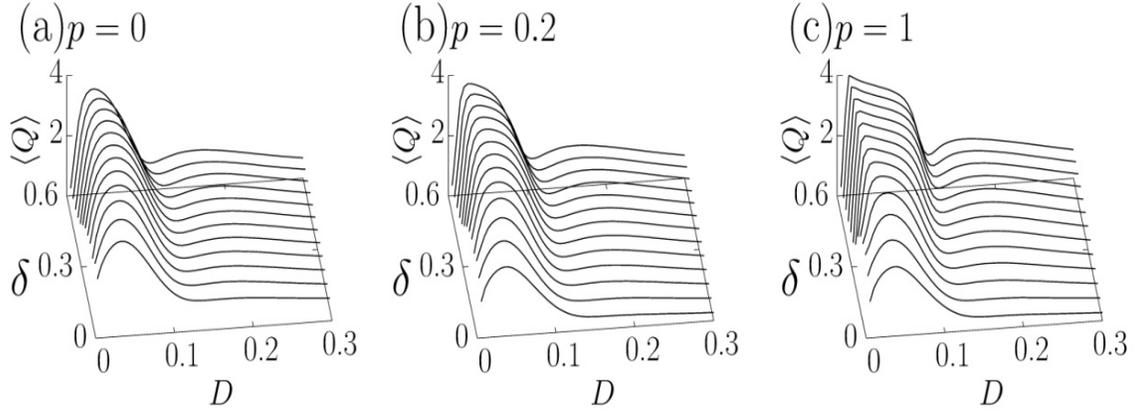

FIG. 14. Variation of $\langle Q \rangle$ of network with $\delta$ and $D$ for (a) $p = 0$ (regular network), (b) $p = 0.2$ (a Watts-Strogatz small-world network) and (c) $p = 1$ (purely random network).

We computed probability distribution of $Q_i$'s for and $1$ and the results are shown in Fig. 13. The distribution of $Q_i$ is relatively very narrow for $D = 0.04$ for each fixed value of $p$. Generally, the width of the distribution of $Q_i$ is relatively small at resonance. The numerically computed average value of $\langle Q \rangle$, is plotted as a function of $\delta$ and $D$ for $p = 0, 0.2$ and $1$ in Fig. 14. The variation $\langle Q \rangle$ with $D$ for each fixed values of $\delta$ and $p$ is smooth and displays resonance behaviour. For $\delta > 0.68$ successive iteration of the state variable $x$ is unbounded.

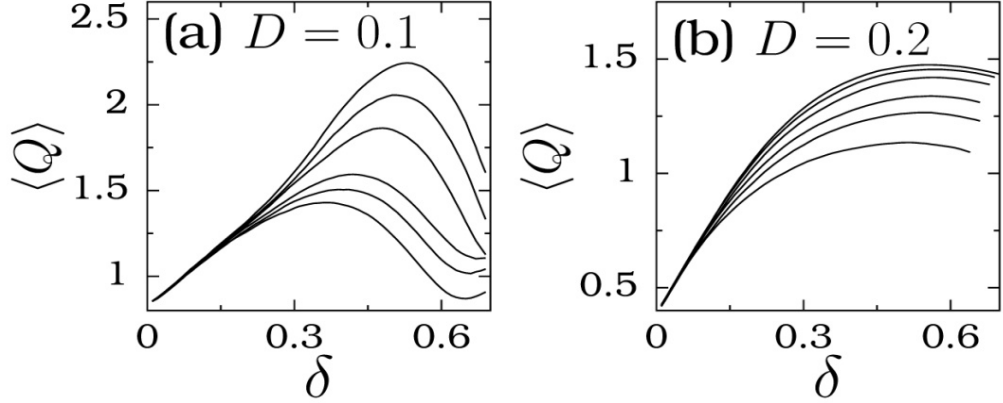

FIG. 15. Dependence of $\langle Q \rangle$ on the coupling strength $\delta$ for (a) $D = 0.1$ and (b) $D = 0.2$ and for several fixed values of $p$ for the network of Bellows maps. The values of $p$ for the curves from bottom to top are 0, 0.2, 0.4, 0.6, 0.8 and 1.

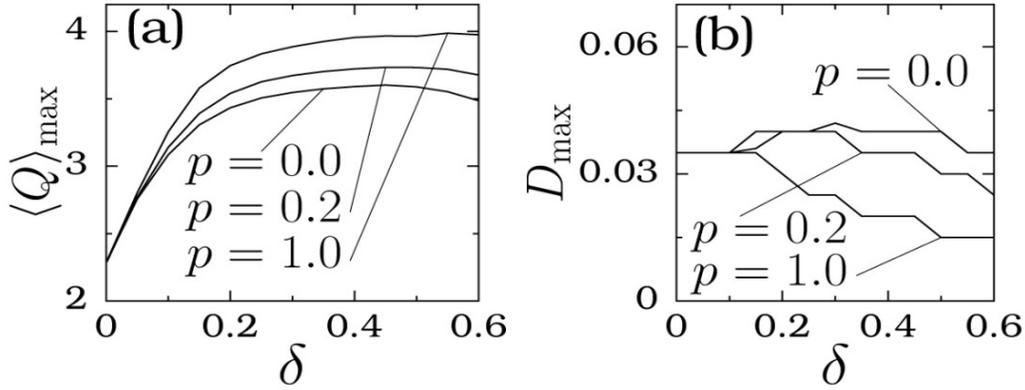

FIG. 16. Variation of $\langle Q \rangle_{max}$ and the corresponding $D_{max}$ as a function of coupling strength $\delta$ for $p = 0, 0.2$ and 1 for the network of Bellows maps.

The influence of the control parameter $\delta$ on $\langle Q \rangle$ and $\langle Q \rangle_{max}$ (the maximum value of $\langle Q \rangle$) is also studied. In Fig. 15 $\langle Q \rangle$ versus $\delta$ is plotted for $D = 0.1$ and $0.2$ for a few fixed values of $p$. For $D = 0.1$ we notice two effects of $p$. Increase in the value of $p$ from $0$ shifts the value of $\delta$ at which $\langle Q \rangle$ becomes maximum. Further, except for values of $\delta$ close to zero the value of $\langle Q \rangle$ for each fixed value of $\delta$ increases with increases in randomness in the connectivity and yields enhanced average response amplitude at the missing fundamental frequency of the input signal. This results is observed for $D = 0.2$ as shown in Fig. 15b. When $D = 0.1$ resonance-like behaviour is exhibited by $\langle Q \rangle$ when $\delta$ is varied. For $D = 0.2$, $\langle Q \rangle$ increases with increases in $\delta$ upto the value $\approx 0.68$ above which successive iterated values of $x$ diverge. Figure 16 presents the dependence of maximum value of $\langle Q \rangle$ on the coupling strength $\delta$ for three values of $\delta$ and reaches a saturation.

## V Conclusions

We have shown the occurrence of ghost-stochastic resonance in two types of network systems. In the unidirectionally coupled network system enhancement of response amplitude at the input multi-frequency signal is found. Each unit displays resonance. Resonance amplitude at the missing frequency varies with the unit number $i$ and found to show saturation. That is, a maximum response can be achieved with a certain optimum number of units in the network. For the distinct units the output signal becomes a sequence of rectangular pulse. $Q_L$ shows distinct types of variation on $D$ and $\delta$. For a fixed $n_f$ the quantity $Q_i$ decreases with increases in the values of $k$. Similar result is found for varying of $n_f$ for fixed $k$. In our analysis best $Q_i$ 's are realized for $k = n_f$. The other combination of $k$ and $n_f$ yields lower $Q_i$. It is noteworthy to point out that realization of enhancement of response amplitude at the missing frequency across the whole network by applying the input signal and noise to the first unit of network is of great useful for weak signal detection, information propagation and generation of signal with frequency missing in the input signal.

In the small-world network $Q_i$ 's are randomly distributed over a range due to the random connectivity. However, $\langle Q \rangle$ exhibited typical resonance when noise intensity is varied. The effect of $\delta$ on $\langle Q \rangle$ is nontrivial as shown in Fig. 15 for different values of $p$. $\langle Q \rangle_{max}$ is found to be maximum for purely random network. The results of our study, in general, can be observed in network of other bistable systems with same connectivity topology.

**Acknowledgment**

S. Rajamani expresses her gratitude to University Grants Commission (UGC), Government of India for financial support in the form of UGC meritorious fellowship.